\documentstyle[epsf,art12,bezier]{article}
\def\si{\quad}
\def\sii{\qquad}

\def\siiii{\qquad\qquad}
\def\b:{\begin{equation}}
\def\e:{\end{equation}}
\def\be:{\begin{eqnarray}}
\def\ee:{\end{eqnarray}}
\def\nn{\nonumber\\}
\def\cl{\centerline}

\def\vo{\vglue 0.3cm}
\def\voo{\vglue 0.5cm}
\def\vi{\vglue 1cm}
\def\vii{\vglue 2cm}

\def\noi{\noindent}

\def\ve{\vfill\eject}
\def\bmin{\begin{minipage}{15cm} }
\def\emin{\end{minipage}}
\newcommand{\lb}[1]{\label{eqn:#1}}
\newcommand{\rf}[1]{\ref{eqn:#1}}
\begin{document}
\baselineskip 20pt
%%%%%%%%%%%%%%%%%%%%%%%%%%%%%%%%%%%%%%%%%%%%%
\vii

\cl{{\Large {\bf Complex Analysis of a Piece of Toda Lattice\footnote{
This work is supported in part by the Grant-in-Aid for general Scientific
Research from the Ministry of Education, Science, Sports and Culture, Japan
(No.06835023), and the Fiscal Year 1996 Fund for Special Research Projects
at Tokyo Metropolitan University.}}}}

\vi
\cl{Satoru SAITO,\ Noriko SAITOH$^\dagger$,\ Hisao KONUMA and Katsuhiko
YOSHIDA}
\vi
\cl{\it Depertment of Physics, Tokyo Metropolitan University,}

\cl{\it  Hachiohji, Tokyo 192-03 Japan }
\vo
\cl{\it $^\dagger$ Department of Applied Mathematics, Yokohama National
 University,}

\cl{\it Hodogaya-ku, Yokohama 240 Japan}
\vglue 8cm
\begin{abstract}
We study a small piece of two dimensional Toda lattice as a complex dynamical
system. In particular the Julia set, which appears when the piece is deformed,
 is shown analytically how it disappears as the system approaches to the
integrable limit.
\end{abstract}

\section{Introduction}

The two dimensional Toda lattice is one of soliton equations which has become
more and more important as a key object in theoretical physics. It was first
formulated by Hirota in 1981\cite{Hirota} as a discrete version of two
continuous time Toda lattice and shown by Miwa\cite{Miwa} its equivalence to
the KP hierarchy. When quasi periodic solutions are substituted, it is nothing
but the identity known as Fay's trisecant formula which characterizes
algebraic curves.

This equation has become known in other fields of physics in the last ten
years. It was shown being satisfied by the string amplitudes in particle
physics\cite{S}. More recently there appeared papers demonstrating unexpected
correlation of this equation with other topics in physics. The transfer matrix
of the solvable lattice model with $A_l$ symmetry, for example, was shown to
satisfy this equation\cite{KLWZ}\cite{Kuniba}. This equation has been also
proven to unify discrete Painlev\'e equations\cite{Ramani}. The connection of
solvable cellular automata to this equation offers another example\cite{TTMS}.

Completely integrable nonlinear systems must play fundamental roles in various
phenomena in physics. It is remarkable that many integrable systems in
different fields are unified into the single equation. We are interested in
clarifying ultimate notion of integrability of the systems. Investigation of
such systems themselves, however, will not reveal all of features of the
systems. The real meaning of integrability will be clarified only in comparison
with nonintegrable systems.

An arbitrary deformation of the two dimensional Toda lattice will destroy
integrability and create chaos. Since the system contains infinite number of
degrees of freedom it is extremely difficult to study analytically the
behaviour of transition from nonintegrable to integrable phases. It should be
recalled that a very little is known about analytical properties of
nonintegrable systems. The main part of the studies of complex dynamical
systems were limited to simple systems with one degree of freedom.

Very recently we pointed out\cite{SS toda30} that a set of lattice points,
which form a parallelogram in the two dimensional lattice space, constitute a
piece of the Toda lattice. We call it a Toda molecule\cite{Toda molecule} since
it is essentially what is intended to be called by this name, but used in a bit
different context in the literature. The remarkable fact is that the small
pieces can be separated from other parts without loosing any properties of the
original Toda lattice.

The purpose of this paper is to study in detail analytical properties of the
smallest piece of Toda molecules. The smallest Toda molecule is a smallest
parallelogram of four lattice points.We will call it a Toda atom for
convenience. Since every Toda molecule preserves properties possessed by the
Toda lattice, we can study analytical properties of the system from the
knowledge of a Toda atom. In the first part of this paper we show that the time
evolution of a Toda atom is described by an iterative M\"obius map. The form
invariance of this map certificates integrability of this system. In the second
part of this paper we will consider a deformation of this piece. Under generic
deformation a chaos will be generated through the time evolution. We are
especially concerned with analytical property of the Julia set as the system
approaches to the integrable map. We will show how the Julia set converges to
the points on the orbit of M\"obius map as a parameter, which interpolates
between integrable and nonintegra

\section{Pieces of Toda Lattice}

In this section we like to show that the two dimensional Toda lattice can be
cut into small pieces without loosing any properties possessed by the original
system. To begin with let us write down the equation which was derived by
Hirota as a discrete version of the two continuous time Toda
lattice\cite{Hirota}:
\be:
&&\alpha\ g_n(l+1,m)g_n(l,m+1)\ +\ \beta\ g_n(l,m)g_n(l+1,m+1)\nn
&&\siiii - (\alpha+\beta)\ g_{n+1}(l+1,m)g_{n-1}(l,m+1)=0,\si \alpha, \beta \in
{\mbox{\boldmath$C$}},\si l,m,n\in \mbox{\boldmath$Z$}.
\lb{HBDE}
\ee:
We called this equation Hirota bilinear difference equation\footnote{This
equation is also called Hirota-Miwa equation in recent literature.} and
abbreviated as HBDE. This is a nonlinear system defined on the three
dimensional lattice space. Our key observation is the following. For a fixed
point of the lattice $(l,m, n)=(\bar l,\bar m,\bar n)$ , we denote by
\mbox{\boldmath$A$} the set of points $(\bar l,\bar m,\bar n), (\bar l+1,\bar
m,\bar n), (\bar l,\bar m+1,\bar n),(\bar l+1,\bar m+1,\bar n) , (\bar l+1,\bar
m,\bar n+1), (\bar l,\bar m+1,\bar n-1)$ . Then if  $g_n(l,m)$ is a solution of
$(\rf{HBDE})$,
\b:
f(l,m,n)=\cases{g_n(l,m),\sii (l,m,n)\in {\mbox{\boldmath$A$}},\cr 0,\siiii\si
{\rm otherwise},\cr}
\lb{Toda atom}
\e:
is also a solution of $(\rf{HBDE})$. This is the smallest piece of the Toda
lattice.

The proof is simple. Because \mbox{\boldmath$A$} is surrounded by zero, every
equation on other pieces is automatically satisfied. The result can be easily
generalized to larger parallelogram prism when it is surrounded by zero. We
call it a Toda molecule according to ref.\cite{Toda molecule}. Then it will be
natural to call $(\rf{Toda atom})$ a Toda atom. If there are many Toda
molecules in the three dimensional lattice space separated by zeros from each
other it is again a solution of  $(\rf{HBDE})$. A slice perpendicular to the
$l$ axis of such example is presented in Fig. 1.

\begin{center}\begin{minipage}{7cm}\unitlength 1mm\begin{picture}(70,75)
\put(0,20){\line(1,0){60}}\put(62,18){\makebox(3,3)[l]{$n$}}
\put(30,5){\line(0,1){55}}\put(28,62){\makebox(3,3)[l]{$m$}}
\put(25,70){\makebox(10,3)[c]{Fig. 1}}
\thicklines
\put(5,55){\circle{1}}\put(10,55){\circle{1}}\put(15,55){\circle{1}}
\put(20,55){\circle{1}}\put(25,55){\circle{1}}
\put(30,55){\circle{1}}\put(35,55){\circle{1}}\put(40,55){\circle{1}}
\put(45,55){\circle{1}}\put(50,55){\circle{1}}\put(55,55){\circle{1}}
\put(5,50){\circle*{1}}\put(10,50){\circle*{1}}\put(15,50){\circle*{1}}
\put(20,50){\circle*{1}}\put(25,50){\circle*{1}}\put(30,50){\circle*{1}}
\put(35,50){\circle*{1}}\put(40,50){\circle{1}}\put(45,50){\circle*{1}}
\put(50,50){\circle*{1}}\put(55,50){\circle*{1}}
\put(5,50){\line(1,0){30}}\put(45,50){\line(1,0){10}}
%% FOLLOWING LINE CANNOT BE BROKEN BEFORE 80 CHAR
\put(5,50){\line(1,-1){5}}\put(10,50){\line(1,-1){5}}
\put(15,50){\line(1,-1){5}}\put(20,50){\line(1,-1){5}}
\put(25,50){\line(1,-1){5}}\put(30,50){\line(1,-1){5}}
\put(35,50){\line(1,-1){5}}\put(45,50){\line(1,-1){5}}
\put(50,50){\line(1,-1){5}}\put(55,50){\line(1,-1){4}}
\put(5,45){\circle*{1}}\put(10,45){\circle*{1}}\put(15,45){\circle*{1}}
\put(20,45){\circle*{1}}\put(25,45){\circle*{1}}\put(30,45){\circle*{1}}
\put(35,45){\circle*{1}}\put(40,45){\circle*{1}}\put(45,45){\circle{1}}
\put(50,45){\circle*{1}}\put(55,45){\circle*{1}}
\put(5,45){\line(1,0){35}}\put(50,45){\line(1,0){7}}
\put(5,40){\circle{1}}\put(10,40){\circle{1}}\put(15,40){\circle{1}}
\put(20,40){\circle{1}}\put(25,40){\circle{1}}
\put(30,40){\circle{1}}\put(35,40){\circle{1}}\put(40,40){\circle{1}}
\put(45,40){\circle{1}}\put(50,40){\circle{1}}\put(55,40){\circle{1}}
\put(5,35){\circle{1}}\put(10,35){\circle{1}}\put(15,35){\circle{1}}
\put(20,35){\circle*{1}}\put(25,35){\circle*{1}}\put(30,35){\circle{1}}
\put(35,35){\circle*{1}}\put(40,35){\circle*{1}}\put(45,35){\circle*{1}}
\put(50,35){\circle{1}}\put(55,35){\circle{1}}
\put(20,35){\line(1,0){5}}\put(35,35){\line(1,0){10}}
\put(20,35){\line(1,-1){28}}\put(25,35){\line(1,-1){28}}
\put(35,35){\line(1,-1){10}}\put(40,35){\line(1,-1){10}}
\put(45,35){\line(1,-1){10}}
\put(5,30){\circle*{1}}\put(10,30){\circle*{1}}\put(15,30){\circle*{1}}
\put(20,30){\circle{1}}\put(25,30){\circle*{1}}
\put(30,30){\circle*{1}}\put(35,30){\circle{1}}\put(40,30){\circle*{1}}
\put(45,30){\circle*{1}}\put(50,30){\circle*{1}}\put(55,30){\circle{1}}
\put(5,30){\line(1,0){10}}\put(25,30){\line(1,0){5}}\put(40,30){\line(1,0){10}}
%% FOLLOWING LINE CANNOT BE BROKEN BEFORE 80 CHAR
\put(5,30){\line(1,-1){5}}\put(10,30){\line(1,-1){5}}
\put(15,30){\line(1,-1){5}}%
\put(5,25){\circle{1}}\put(10,25){\circle*{1}}\put(15,25){\circle*{1}}
\put(20,25){\circle*{1}}\put(25,25){\circle{1}}\put(30,25){\circle*{1}}
\put(35,25){\circle*{1}}\put(40,25){\circle{1}}\put(45,25){\circle*{1}}
\put(50,25){\circle*{1}}\put(55,25){\circle*{1}}
\put(10,25){\line(1,0){10}}\put(30,25){\line(1,0){5}}
\put(45,25){\line(1,0){10}}
\put(5,20){\circle{1}}\put(10,20){\circle{1}}\put(15,20){\circle{1}}
\put(20,20){\circle{1}}\put(25,20){\circle{1}}
\put(30,20){\circle{1}}\put(35,20){\circle*{1}}\put(40,20){\circle*{1}}
\put(45,20){\circle{1}}\put(50,20){\circle{1}}\put(55,20){\circle{1}}
\put(35,20){\line(1,0){5}}
\put(5,15){\circle{1}}\put(10,15){\circle{1}}\put(15,15){\circle*{1}}
\put(20,15){\circle*{1}}\put(25,15){\circle{1}}
\put(30,15){\circle{1}}\put(35,15){\circle{1}}\put(40,15){\circle*{1}}
\put(45,15){\circle*{1}}\put(50,15){\circle{1}}\put(55,15){\circle{1}}
\put(15,15){\line(1,0){5}}\put(40,15){\line(1,0){5}}\put(15,15){\line(1,-1){5}}
\put(20,15){\line(1,-1){5}}
\put(5,10){\circle{1}}\put(10,10){\circle{1}}\put(15,10){\circle{1}}
\put(20,10){\circle*{1}}\put(25,10){\circle*{1}}
\put(30,10){\circle{1}}\put(35,10){\circle{1}}\put(40,10){\circle{1}}
\put(45,10){\circle*{1}}\put(50,10){\circle*{1}}\put(55,10){\circle{1}}
\put(20,10){\line(1,0){5}}\put(45,10){\line(1,0){5}}
\end{picture}\end{minipage}\end{center}

For an illustration let us consider the one soliton state localized on the
smallest parallelogram specified by $(m,n)=(0,0), (0,1), (1,-1), (1,0)$ on the
$(m,n)$ lattice plane, but allowed to range all integers along $l$. Now we
recall that in the usual lattice space the one soliton solution is given
by\cite{Miwa}\cite{SS3}
\b:
f^{1sol}(l,m,n)=\prod_j(1-az_j)^{-k_j}+\prod_j(1-bz_j)^{-k_j}.
\lb{Miwa1soliton}
\e:
Here $a,b$ are arbitrary constants and $\{z_j\}$ are parameters which determine
velocity of the soliton. $\{k_j\}$ are variables taking values on integers. We
can choose any three among $\{k_j\}$ to relate them to our variables $(l,m,n)$.
Let $k_1, k_2, k_3$ be such three and relate them according to
\b:
k_1=m+n-{1\over 2},\si k_2=-m-{1\over 2},\si k_3=l-n-{1\over 2}.
\e:
Writing $(\rf{Miwa1soliton})$ explicitly we find
\be:
f^{1sol}(l,0,0)&=&A(1-az_3)^{-l}+B(1-bz_3)^{-l}\nn
f^{1sol}(l,1,0)&=&A{1-az_2\over 1-az_1}(1-az_3)^{-l}+B{1-bz_2\over
1-bz_1}(1-bz_3)^{-l}\nn
f^{1sol}(l,0,1)&=&A{1-az_3\over 1-az_1}(1-az_3)^{-l}+B{1-bz_3\over
1-bz_1}(1-bz_3)^{-l}\nn
f^{1sol}(l,1,-1)&=&A{1-az_2\over 1-az_3}(1-az_3)^{-l}+B{1-bz_2\over
1-bz_3}(1-bz_3)^{-l}
\lb{1soliton}
\ee:
where
$$
A:=\sqrt{(1-az_1)(1-az_2)(1-az_3)},\si B:=\sqrt{(1-bz_1)(1-bz_2)(1-bz_3)}.
$$
We see from $(\rf{1soliton})$ that all points belonging to the same piece
behave similarly. The parameters are related to $\alpha,\ \beta$ of
$(\rf{HBDE})$ by
$$
\alpha=z_1(z_2-z_3),\si \beta=z_2(z_3-z_1),
$$
for $(\rf{1soliton})$ to satisfy HBDE. If we define the amplitude
$\varphi_{mn}(l)$ by
\b:
\varphi_{mn}(l):={f(l+1,m,n)f(l-1,m,n)\over f^2(l,,m,n)}\ -\ 1
\lb{amplitude}
\e:
it behaves as
\b:
\varphi_{00}^{1sol}(l)={\sinh^2p\over \cosh^2 (pl+\chi)},\si p:={1\over
2}\ln{1-az_3\over 1-bz_3},\si \chi:={1\over 2}\ln{B\over A}.
\lb{soliton peak}
\e:
This represents a localized peak along the $l$ axis. The other amplitudes
$\varphi_{mn}(l)$ behave almost the same but different by the values of the
phase $\chi$.
%%%%%%%%%%%%%%%%%%%%%%%%%%%%%%%%%%%%

If we consider an evolution of the system in variable $l$, a Toda atom is
composed of four lattice points. Since there is only one equation of motion
$(\rf{HBDE})$, they are not independent variables. Three of them can be chosen
as we like leaving one to be determined by the equation. Let $z_l$ be
$f(l,0,0)$. The other three could be either dependent or independent of $z_l$.
If they are independent of $z_l$, the equation of motion is linear in $z_l$. On
the other hand if they do depend on $z_l$ they are allowed at most linear in
$z_l$, for the equation to remain Hirota bilinear form. Namely we can write
\b:
f(l,m,n)=A_{m,n}z_l+B_{m,n},\sii (m,n)=(0,0),(1,0),(0,1),(1,-1),
\lb{linear transf}
\e:
with $A_{0,0}=1,\ B_{0,0}=0$. Upon substituting them together into HBDE, it is
easy to see that we obtain an equation of the form
\b:
z_{l+1}={Az_l+B\over Cz_l+D}.
\lb{general Moebius}
\e:
$$
A=-\beta B_{1,0}+(\alpha+\beta)B_{0,1}A_{1,-1},\si
B=(\alpha+\beta)B_{0,1}B_{1,-1},
$$
$$
C=(\alpha+\beta)(A_{1,0}-A_{0,1}A_{1,-1}),\si D=\alpha
B_{1,0}-(\alpha+\beta)A_{0,1}B_{1,-1}.
$$
$(\rf{general Moebius})$ is a M\"obius map. Thererfore the map is integrable.

If we remember that HBDE is invariant under the transformation of
$f(l,m,n)\rightarrow e^{al+bm+cn}f(l,m,n)$, the one soliton solution
$(\rf{1soliton})$ offers an example of $(\rf{linear transf})$.

The solution of $(\rf{general Moebius})$ can be obtained as follows. A M\"obius
map has three fixed points. By an appropriate transformation : $z_l\rightarrow
\phi\circ z_l\circ \phi^{-1}$, one of the fixed points can be transformed into
0. After the transformation the map will have the form
\b:
z_{l+1}=\mu{z_l\over 1+\nu z_l}.
\lb{ILM}
\e:
$(\rf{ILM})$ is easily solved for an arbitrary initial value $z_0$ to get
\b:
z_l={\mu^l z_0\over 1+\nu{1-\mu^l\over 1-\mu}z_0}.
\lb{Moebius}
\e:
Applying to this the inverse transformation : $z_l\rightarrow \phi^{-1}\circ
z_l\circ \phi$ , the general solution to $(\rf{general Moebius})$ is obtained.
We call the map $(\rf{ILM})$ the integrable logistic map (ILM). The meaning of
this name will become clear later.

We notice that $(\rf{ILM})$ corresponds to the case in which one of the lattice
point is fixed constant and the other three points behave the same:
\be:
f^{ILM}(l,0,0)&=&f^{ILM}(l,0,1)=f^{ILM}(l,1,-1)=:z_l,\nn
f^{ILM}(l,1,0)&=&{\mu-1\over\nu},\si \mu=-{\beta\over\alpha}.
\lb{simple case}
\ee:

How does the amplitude look like in this case? To see it we substitute
$(\rf{Moebius})$ into $(\rf{amplitude})$ and get
\b:
\varphi^{ILM}={\sinh^2p\over\cosh^2(pl+\chi)-\cosh^2 p},\si p:={1\over 2}\ln\mu
,\si \chi:={1\over 2}\ln{\mu z_0\over 1-\mu+\nu z_0}.
\e:
The similarity of this result to the one soliton solution $(\rf{soliton peak})$
must be apparent.

We may further simplify the equation by
\b:
f^{lin}(l,0,0)=:z_l,\si f^{lin}(l,1,-1)=f^{lin}(l,1,0)=1-{1\over\mu},\si
f^{lin}(l,0,1)=c (c: {\rm const}).
\e:
The map turns to be linear
\b:
z_{l+1}=\mu z_l +(1-\mu)c
\e:
and yields the solution
\b:
z_l=\mu^l (z_0-c)+c.
\e:
The corresponding amplitude is
\b:
\varphi^{lin}(l)={\sinh^2p\over\cosh^2(pl+\chi)},\si p={1\over 2}\ln\mu,\si
\chi={1\over 2}\ln{z_0-c\over c}.
\e:
which is again the form of $(\rf{soliton peak})$.

\section{Generalized Logistic Map}

As we have learned in the preceeding section the smallest piece of Toda lattice
already possesses useful informations of the integrable dynamical systems. In
this section we study a deformation of the Toda atom. There could be many
different ways of deformation, some of which preserve integrability and some
others destroy it. Since we are interested in studying the transition between
integrable and nonintegrable maps, we must break integrability of the Toda
atom.

For this purpose we recall that the Toda molecules have a characteristic form
as seen in Fig. 1. Their cross sections in the $(m,n)$ plane are parallelogram
declined to the same direction. It owes to the property of the Toda atom
defined in $(\rf{Toda atom})$. The very reason of this asymmetry comes from the
asymmetry\footnote{If we had chosen other set of variables, HBDE looked more
symmetric and the corresponding Toda atom could be either cubic or
octahedron\cite{DYB}. We have used asymmetric variables such that deformations
can be discussed.} of HBDE under the exchange of $l$ and $m$ as seen in
$(\rf{HBDE})$. The equation in which the role of $l$ and $m$ in $(\rf{HBDE})$
are exchanged is also integrable. In fact we could start from it without
changing none of the results.

 From this argument we are tempted to consider the following deformation of
HBDE.
\be:
&&\alpha\ g_n(l+1,m)g_n(l,m+1)+\beta\ g_n(l,m)g_n(l+1,m+1)\nn
&&\sii -\ (\alpha+\beta)\ \left[(1-\gamma)\delta g_{n+1}(l+1,m)+\gamma
g_{n+1}(l,m+1)\right]\nn
&&\siiii\sii \times\ \left[(1-\gamma')\delta' g_{n-1}(l,m+1)+\gamma'
g_{n-1}(l+1,m)\right]=0.
\lb{deformed HBDE}
\ee:
We notice that this equation is integrable when $\gamma=\gamma'=0$ and
$\delta\delta'=1$ ,or $\gamma=\gamma'=1$. Integrability of other cases is not
known at this point. Moreover we are not able to separate some small part of
lattice independently from the rest as it was done to get a Toda atom.
Nevertheless it is worthwhile to study $(\rf{deformed HBDE})$ defined on a
portion of the lattice shown in Fig. 2a.

%%%%%%%%%%%%%%%%%%%%%%%%%%%%%%%%%%%%%%%%%%%%%%%%%%%%%%%
\begin{center}\begin{minipage}{6cm}\unitlength 1mm\begin{picture}(60,50)
%% FOLLOWING LINE CANNOT BE BROKEN BEFORE 80 CHAR
\put(5,5){\circle{2}}\put(15,5){\circle{2}}\put(25,5){\circle{2}}
\put(35,5){\circle{2}}\put(45,5){\circle{2}}
%% FOLLOWING LINE CANNOT BE BROKEN BEFORE 80 CHAR
\put(5,15){\circle{2}}\put(15,15){\circle*{2}}\put(25,15){\circle*{2}}
\put(35,15){\circle*{2}}\put(45,15){\circle{2}}
%% FOLLOWING LINE CANNOT BE BROKEN BEFORE 80 CHAR
\put(5,25){\circle{2}}\put(15,25){\circle*{2}}\put(25,25){\circle*{2}}
\put(35,25){\circle*{2}}\put(45,25){\circle{2}}
%% FOLLOWING LINE CANNOT BE BROKEN BEFORE 80 CHAR
\put(5,35){\circle{2}}\put(15,35){\circle{2}}\put(25,35){\circle{2}}
\put(35,35){\circle{2}}\put(45,35){\circle{2}}
\thicklines
%% FOLLOWING LINE CANNOT BE BROKEN BEFORE 80 CHAR
\put(15,25){\line(1,0){20}}\put(15,25){\line(0,-1){10}}
\put(15,15){\line(1,0){20}}\put(35,25){\line(0,-1){10}}
\bezier{15}(25,25)(30,20)(34,16)\bezier{15}(15,25)(20,20)(24,16)
\put(20,45){\makebox(10,3)[c]{Fig. 2a}}
\end{picture}\end{minipage}
%%%%%%%%%%%%%%%%%%%%%%%%%%%%%%%%%%%%%%%%%%%%%%%%%%%%%%%%%%%%
\begin{minipage}{6cm}\unitlength 1mm\begin{picture}(60,50)
%% FOLLOWING LINE CANNOT BE BROKEN BEFORE 80 CHAR
\put(5,5){\circle{2}}\put(15,5){\circle{2}}\put(25,5){\circle{2}}
\put(35,5){\circle{2}}\put(45,5){\circle{2}}
%% FOLLOWING LINE CANNOT BE BROKEN BEFORE 80 CHAR
\put(5,15){\circle{2}}\put(15,15){\circle{2}}\put(25,15){\circle*{2}}
\put(35,15){\circle*{2}}\put(45,15){\circle{2}}
%% FOLLOWING LINE CANNOT BE BROKEN BEFORE 80 CHAR
\put(5,25){\circle{2}}\put(15,25){\circle*{2}}\put(25,25){\circle*{2}}
\put(35,25){\circle*{2}}\put(45,25){\circle{2}}
%% FOLLOWING LINE CANNOT BE BROKEN BEFORE 80 CHAR
\put(5,35){\circle{2}}\put(15,35){\circle{2}}\put(25,35){\circle{2}}
\put(35,35){\circle{2}}\put(45,35){\circle{2}}
\thicklines
%% FOLLOWING LINE CANNOT BE BROKEN BEFORE 80 CHAR
\put(15,25){\line(1,0){20}}\put(15,25){\line(1,-1){10}}
\put(35,25){\line(0,-1){10}}\put(25,15){\line(1,0){10}}
\bezier{15}(25,25)(30,20)(34,16)
\put(20,45){\makebox(10,3)[c]{Fig. 2b}}
\end{picture}\end{minipage}\end{center}
%%%%%%%%%%%%%%%%%%%%%%%%%%%%%%%%%%%%%%%%%%%%%%%%%%%%%%%%%%%%
\voo
In order to proceed further we have to specify the model so that we can study
analytical properties of the map explicitly. We will consider, in the following
discussion, the map given by $(\rf{ILM})$ and its deformation. We also restrict
our argument to the case of $\gamma'=0$, $\delta=\delta'^{-1}={\nu\over\mu}$ in
$(\rf{deformed HBDE})$ for simplicity and define (Fig. 2b)
$$
f^{GLM}(l,0,0)=f^{GLM}(l,0,1)=f^{GLM}(l,1,-1)=f^{GLM}(l,1,1)=:z_l,
$$
\b:
f^{GLM}(l,1,0)={\mu -1\over\nu}\ (\mu: {\rm const}).
\e:
The dynamics of this model is described by the map
\b:
z_{l+1}=f(z_l):=\mu{z_l(1-\gamma z_l)\over 1+\nu(1-\gamma)z_l};\si z_l \in
\mbox{\boldmath$C$},\si l\in \mbox{\boldmath$Z$}.
\lb{GLM}
\e:
We call this map a generalized logistic map (GLM). Some properties are listed
below :

\begin{enumerate}
\item
When $\gamma=1$, the map becomes the ordinary logistic map studied in the
literature extensively.
\item
GLM becomes the logistic equation for all values of the parameters $\gamma,\mu
,\nu$ when the continuous limit of the variable $l$ is taken. To show it let us
introduce new variables $u$ and new parameters $a$ and $h$ by
\b:
u(l):={\nu+\gamma\mu-\gamma\nu\over\mu-1}z_l,\sii ah:=\mu-1.
\e:
We replace $z_{l+1}$ by $z_{l+h}$ and take the limit $h\rightarrow 0$. We will
find that $(\rf{GLM})$ reduces to the logistic equation:
\b:
{du\over dl}=au(1-u).
\lb{logistic equation}
\e:
\item
GLM includes $(\rf{ILM})$ as the special case with $\gamma=0$. This explaines
the name of ILM used for $(\rf{ILM})$.
\item
GLM generates Julia set as long as $\gamma\ne 0$. Hence it is not integrable
except for $\gamma=0$. This will be discussed later.
\end{enumerate}

The most important feature of GLM is that it interpolates nonintegrable map to
integrable map in the limit of continuous deformation. This fact enables us to
study analytically the transition between two phases. The problem we concern in
what follows is the analytical properties of the map $(\rf{GLM})$. To proceed
further it is more convenient to convert the map $(\rf{GLM})$ into the standard
form of rational map of degree 2:
\b:
F(z)=\phi\circ f\circ \phi^{-1}(z)={z(z+\lambda)\over 1+\lambda' z}\
e^{i\theta}\lb{F(z)}
\e:
by the M\"obius transformation
\b:
\phi(x)={(1-\mu)x\over(\nu\gamma-\nu-\mu\gamma)x+(\nu\gamma-\nu-\gamma)\mu
e^{-i\theta}}.
\e:
where
\b:
\lambda=\mu
e^{-i\theta},\sii\lambda'={\nu\gamma-\nu-2\mu\gamma+\mu^2\gamma\over
(\nu\gamma-\nu-\gamma)\mu} e^{i\theta}.
\e:
The corresponding integrable map turns to be the following case
\b:
F(z)=\mu z=\lambda e^{i\theta}z,\si {\rm if}\si \gamma=0\si {\rm i.e.} \si
\lambda\lambda'=1.
\lb{linear map}
\e:

The main feature of a dynamical system is determined by the nature of fixed
points of the map. Namely the multiplier $\Lambda$ at fixed point $a$ of the
map $\varphi(z)$ is defined by the derivative of the map at $a$:
\b:
\Lambda:=\left.{d\varphi(z)\over dz}\right|_{z=a}.
\e:
The fixed point $a$ is an attractor of the map if $|\Lambda|<1$, a repeller if
$|\Lambda|>1$, and neutral if $|\Lambda|=1$.

In the case of GLM, the fixed points are easily found as
\b:
0,\si p=-\lambda-{1-\lambda\lambda'\over\lambda'-e^{i\theta}},\si \infty.
\e:
The corresponding multipliers are
\b:
\Lambda_0=\lambda e^{i\theta},\si \Lambda_p={2-\lambda
e^{i\theta}-\lambda'e^{-i\theta}\over 1-\lambda\lambda'},\si
\Lambda_\infty=\lambda' e^{-i\theta}.
\e:

In the integrable limit $\lambda\lambda'\rightarrow 1$, we observe the
following characteristic features:
\begin{enumerate}
\item
Since
\b:
|\Lambda_0\Lambda_\infty|\rightarrow 1,
\e:
the map converges either to $0$ or to $\infty$ depending on $|\lambda|=|\mu|<1$
or $>1$.
\item
The fixed point $p$ approaches to $-\lambda$ and it turns to a super repeller
\b:
|\Lambda_p|\ \rightarrow\ \infty.
\e:
\end{enumerate}

\section{Julia Sets}

In the complex dynamical systems, chaos appears from a Julia set.
Given the map $f(z)$ on a Riemann sphere
$\bar{\mbox{\boldmath $C$}}=\mbox{\boldmath $C$}\cup \{ \infty \}$,
the Riemann sphere is devided into two parts
depending on whether the orbits converge or not.
A set of initial values whose orbits,together with their neiborhood, converge
is called Fatou set $F(f)$.
On the other hand, a set which does not is called Julia set $J(f)$.
This definition leads to the fact
that the Julia set does not contain any attractive periodic cycle.
In this sense the orbit in Julia set is chaotic.

By definiton, Fatou set and Julia set are invariant of the map,
that is \[ f(F)=f^{-1}(F)=F\, ,\: f(J)=f^{-1}(J)=J. \]
It is easy to understand that attractive fixed points belong to Fatou set.
Contrary it is known that repulsive fixed points belong to Julia set
\cite{chaos}.
Then we can compute Julia set by
inversely mapping a repulsive fixed point as an initial value.
We show some of their examples in Fig. 3 for the map of $(\rf{F(z)})$.

The Julia set does not exist if the map is completely integrable. Integrable
maps converge to orbits predictable for any given initial values. Conversely if
there exists an orbit not predictable for some initial values, the map is not
integrable. Therefore a Julia set appears in nonintegrable maps, but not in
integrable maps.

In our standard map of degree 2 given by $(\rf{F(z)})$, a Julia set is known to
exists except for at the integrable point $\lambda\lambda'=1$. We like to know
how it disappears from the complex plane of the variable when the parameters
approach to the limit $\lambda\lambda'\ \rightarrow\ 1$. We have given in
\cite{SSSY} an argument about this problem for some limited range of
parameters. The purpose of this section is to present another argument which
should supplement our previous one.

The inverse map of $(\rf{F(z)})$ is easily obtained as
\b:
z_l=F^{-1}(z_{l+1})={1\over 2}(\rho  z_{l+1}-\lambda)\pm{1\over 2}\sqrt{(\rho
z_{l+1}+\lambda)^2+4z_{l+1}e^{-i\theta}(1-\lambda\lambda')},
\lb{inverse map}
\e:
where we defined
\b:
\rho:=\lambda'e^{-i\theta}.
\e:
 From this expression it is apparent that the inverse map is not unique but
double valued at every step. As we pointed out in above the inverse map
generates points of the Julia set if it starts from a point on the Julia set.
Substituting one value of the Julia set into $(\rf{inverse map})$, we get two
points every time. After $n$ steps the number of points of the Julia set
increases as many as $2^{n+1}-1$. This explains the nature of the Julia set.
Some of the points could be those of periodic maps. They must be subtracted
from the number.

In the integrable limit $\lambda\lambda'\ \rightarrow\ 1$ the inverse map
$(\rf{inverse map})$ is still double valued. They are
\b:
z_l=\cases{\rho z_{l+1}\cr -\lambda\cr}.
\lb{ILM set}
\e:
We notice that the second solution does not depend on $z_{l+1}$, hence is the
same at every step of the map. For $(\rf{ILM set})$ to generate the Julia set
we must start from a repulsive fixed point. When $|\lambda|>1$, the origin is
such a point. Thence we find from $(\rf{ILM set})$ the `Julia
set'\footnote{This set does not possess  properties expected for the ordinary
Julia set. We call it `Julia set' only in the sense that it is generated by the
inverse map starting from a repeller.}
\b:
J^{ILM}=\left\{\left. -\rho^n\lambda\ \right|\ n\in
{\mbox{\boldmath$N$}}\right\}\e:
for the integrable map. The number of the `Julia set' increases proportional to
the number of the steps $n$. Moreover the element of $J^{ILM}$ is equal to
$-\mu^{-n}\lambda$, which is nothing but the solution exactly expected from the
map $(\rf{linear map})$, if it started from $-\lambda$.

The next problem we concern is to explore how the Julia set of GLM turns into
those points of $(\rf{ILM set})$ in the limit $\lambda\lambda'\ \rightarrow\
1$. Since we are interested in the transition from a nonintegrable map to the
integrable map, we are to consider small values of $|\lambda\lambda'-1|$. The
inverse map $(\rf{inverse map})$ can be rewritten as
\b:
F^{-1}(z)=\left\{\matrix{\rho z\cr -\lambda\cr}\right\}\pm E(z)
\lb{F^-1}
\e:
where we put
\b:
E(z):={1\over 2}(\rho z+\lambda)\left(\sqrt{1-{4z\epsilon
e^{-i\theta}\over(\rho z+\lambda)^2}}\ -\
1\right),\sii\epsilon:=\lambda\lambda'-1.
\lb{E(z)}
\e:
Note that $E(z)$ vanishes for small values of $\epsilon$. To see the behaviour
of $E(z)$ for small $\epsilon$ we first observe the inequality which is true
for all $\epsilon$:
\b:
|E(z)|< 3\sqrt{|z\epsilon|},\sii^\forall\epsilon\in {\mbox{\boldmath$C$}}.
\lb{|E(z)|<}
\e:
The proof of this inequality owes to the following facts.
\begin{enumerate}
\item
If $|w|<1$,
\be:
\left|{1\over \sqrt w}\left(\sqrt{1-w}-1\right)\right|&=&{1\over |\sqrt
w|}\left|1-\sqrt{1-w}\right| \le {1\over |\sqrt
w|}\left(1-\sqrt{1-|w|}\right)\nn
&\le&{1\over |\sqrt w|}\left(1-(1-|w|)\right)=|\sqrt w|\le 1.
\ee:
\item
If $|w|>1$,
\be:
\left|{1\over \sqrt w}\left(\sqrt{1-w}-1\right)\right|&=&\left|\sqrt{{1\over
w}-1}-\sqrt{{1\over w}}\right|< 3.
\ee:
\end{enumerate}

Substituting
\b:
w:={4z\epsilon e^{-i\theta}\over (\rho z+\lambda)^2},
\lb{w}
\e:
into $E(z)$ of $(\rf{E(z)})$, we can write
\b:
|E(z)|=\sqrt{|z\epsilon|}\left|{1\over \sqrt
w}\left(\sqrt{1-w}-1\right)\right|,\e:
from which $(\rf{|E(z)|<})$ follows.

We can perform the inverse map $(\rf{F^-1})$ iteratively. Let us denote the map
$(\rf{F^-1})$ as
\b:
A(z):=\rho z+E(z),\sii B(z):=-\lambda-E(z).
\e:
Then the second map becomes
\b:
F^{-2}(z)=\cases{A\left(F^{-1}(z)\right)\cr
B\left(F^{-1}(z)\right)\cr}=\cases{A\circ A(z)\cr A\circ B(z)\cr B\circ A(z)
\cr B\circ B(z)\cr}.
\e:
After $n$ steps we obtain
\b:
F^{-n}(z)=\left\{\left. A^{\nu_1}\circ B^{\nu_2}\circ A^{\nu_3}\circ\cdots\circ
B^{\nu_n}(z)\right|\ \nu_1+\nu_2+\cdots+\nu_n=n\right\}.
\e:

If we had started from the repeller the above maps have produced the Julia set
of GLM. In the following we consider the case $|\lambda|>1,\ |\lambda'|<1$, so
that the origin is a repulsive fixed point and the infinity is an attractive
fixed point. Since $E(0)=0$ the origin is mapped to
\b:
A(0)=0,\sii B(0)=-\lambda
\e:
by the first iteration. The second iteration yields
\b:
A^2(0)=0,\ A\circ B(0)=-\rho\lambda+E(-\lambda),\ B\circ A(0)=-\lambda,\
B^2(0)=-\lambda-E(-\lambda).
\lb{2 iteration}
\e:
We notice that since $E(-\lambda)$ is the order of $\epsilon$ from
$(\rf{|E(z)|<})$, all of the points after the second iteration are in the
neighbourhood of $J^{ILM}$. Proceeding similarly we obtain the Julia set as
follows:
\b:
J^{GLM}=\left\{\ \left. A^{\nu_1}\circ B^{\nu_2}\circ A^{\nu_3}\circ
\cdots\circ B^{\nu_\infty}(0)\right|\ \nu_1,\nu_2,\cdots \in
{\mbox{\boldmath$N$}} \right\}.
\lb{J^GLM}
\e:

We remark some important properties which result from this expression.
\begin{enumerate}
\item
The invariance of the Julia set under the map.

It is obvious from $(\rf{J^GLM})$ that
\b:
J^{GLM}=A\left(J^{GLM}\right)\cup
B\left(J^{GLM}\right)=F^{-1}\left(J^{GLM}\right).
\e:
\item
An element of the form $B\circ X$ for any $X\in J^{GLM}$ belongs to the
neibourhood of $-\lambda$, as seen from
\b:
B\circ X=-\lambda-E(X).
\e:
\item
An element of the form $A^s\circ B\circ X$ maps $B\circ X$ to the neighbourhood
of $-\rho^s\lambda$.

In fact after applying $A$'s $s$ times we get
\be:
A^s(B\circ X)&=&A^{s-1}(\rho B\circ X+E(B\circ X))\nn
&=&A^{s-2}\left(\rho^2 B\circ X+\rho E(B\circ X)+E(A\circ B\circ X)\right)\nn
&=&\rho^s B\circ X
   +\sum_{k=0}^{s-1}\rho^k E\left(A^{s-k-1}\circ B\circ X\right)\nn
&=&-\rho^s\lambda-\rho^s E(X)
   +\sum_{k=0}^{s-1}\rho^k E\left(A^{s-k-1}\circ B\circ X\right).
\lb{A^s(BX)}
\ee:
\end{enumerate}
Since every element of $J^{GLM}$, beside 0, is either the form of $B\circ X$ or
$A^s\circ B\circ X$, we conclude that every element of $J^{GLM}$ is in the
neighbourhood of $J^{ILM}$.

We now proceed to show that $J^{GLM}$ approaches uniformly to $J^{ILM}$ as
$\epsilon$ goes to 0. Since the infinity is an attractive fixed point the Julia
set must be in a finite region of the complex plane. We assume that they are
inside of the disc of radius $R$, {\it i.e.}, $|z|< R,\ ^\forall z\in J^{GLM}$.
%We also notice that the denominator $\rho z+\lambda$ of $(\rf{w})$ does not
% vanish when $z\in J^{GLM}$. Hence the conditions used to derive $(\rf{bound
% of E})$ are fulfilled when $z$'s are points on $J^{GLM}$. Finally we get
Therefore we can bound $|E(z)|$ by
\b:
|E(z)|< 3\sqrt R\sqrt{|\epsilon|},\sii z\in J^{GLM}.
\e:

The summation of $(\rf{A^s(BX)})$ can be estimated as
\b:
\sum_{k=0}^{s-1}\left|\rho^kE\left(A^{s-k-1}BX\right)\right|<
\left(\sum_{k=0}^{s-1}|\rho|^k\right)3\sqrt R\sqrt{|\epsilon|}={1-|\rho|^s\over
1-|\rho|}3\sqrt R\sqrt{|\epsilon|},
\e:
which vanishes as $|\epsilon|$ approaches to 0 for all integer $s$ because we
assume $|\rho|^s=|\lambda'|^s < 1$. This proves that all points in $J^{GLM}$
approach to $J^{ILM}$ uniformly in the integrable limit.

In the above we considered the case of $|\lambda|>1,|\lambda'|<1$.
The other case $|\lambda|>1,|\lambda'|<1$ can be also treated similarly
if $z_l$ is transformed into $w_l=1/z_l$ in (\rf{F(z)}).
Since this transformation is equivalent to the exchange of the role of
$\lambda$ and $\lambda'$ (and replacement of $\theta$ by $-\theta$ )
in (\rf{F(z)}), we can replay on the $w$-plane the same argument to the above.

We conclude this paper by showing pictures which represent the convergence of
the Julia set to the points of iterative maps of the integrable system. The
parameters of the map are fixed at $\lambda=4$ and $\theta=0.03\pi$. Under the
choice of these parameters, $\lambda'$, hence $\epsilon$, can be changed
freely. In the integrable limit $\epsilon=0$, $J^{ILM}=\left\{-4{1\over
4^n}e^{-i0.03\pi n},\ n=0,1,2,\cdots\right\}$.It is shown in Fig. 3c. As
$\epsilon$ differs from 0 the Julia set expand from these points as seen in
other pictures. The real and imaginary axes are not drawn except in Fig. 3a, so
that the points in the neighbourhoods of $z=-4$ and 0 are visible in other
figures.
\voo
\noi
%%%%%%%%%%%%%%%%%%%%%%%%%%%%%%%%%%%%%%%%%%%%%%%%%%%%%%%%%%%%%%%%%%%%%%%%%%
\begin{picture}(80,80)
\put(0,0){\epsfxsize=8cm\epsfbox{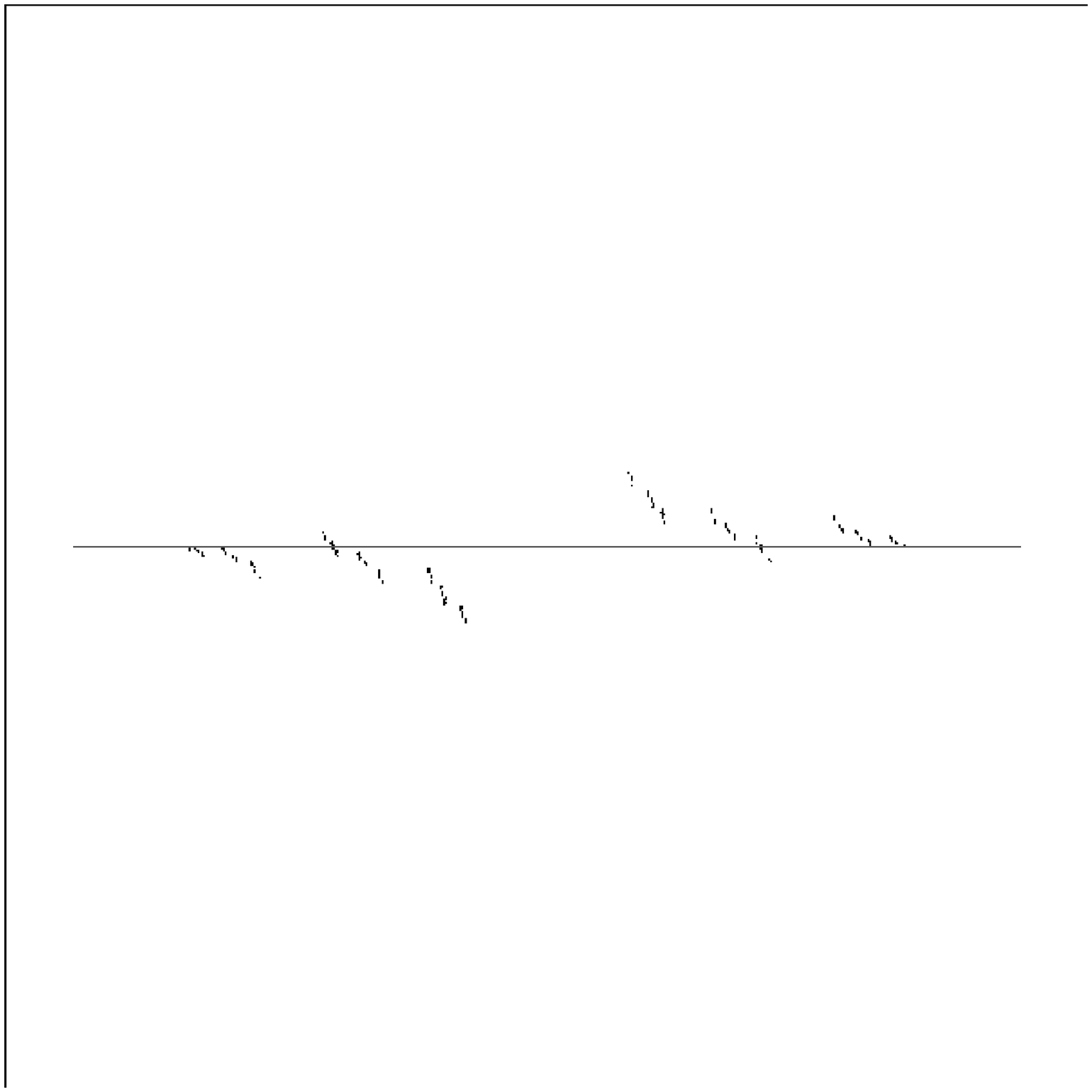}}
\put(35,70){\makebox(10,3)[c]{Fig. 3a}}
\put(10,35){\makebox(4,4)[c]{$- 4$}}
\put(67,35){\makebox(4,4)[c]{0}}
\put(35,5){\makebox(10,3)[c]{$\epsilon=-1$}}
\end{picture}
\begin{picture}(80,80)
\put(0,0){\epsfxsize=8cm\epsfbox{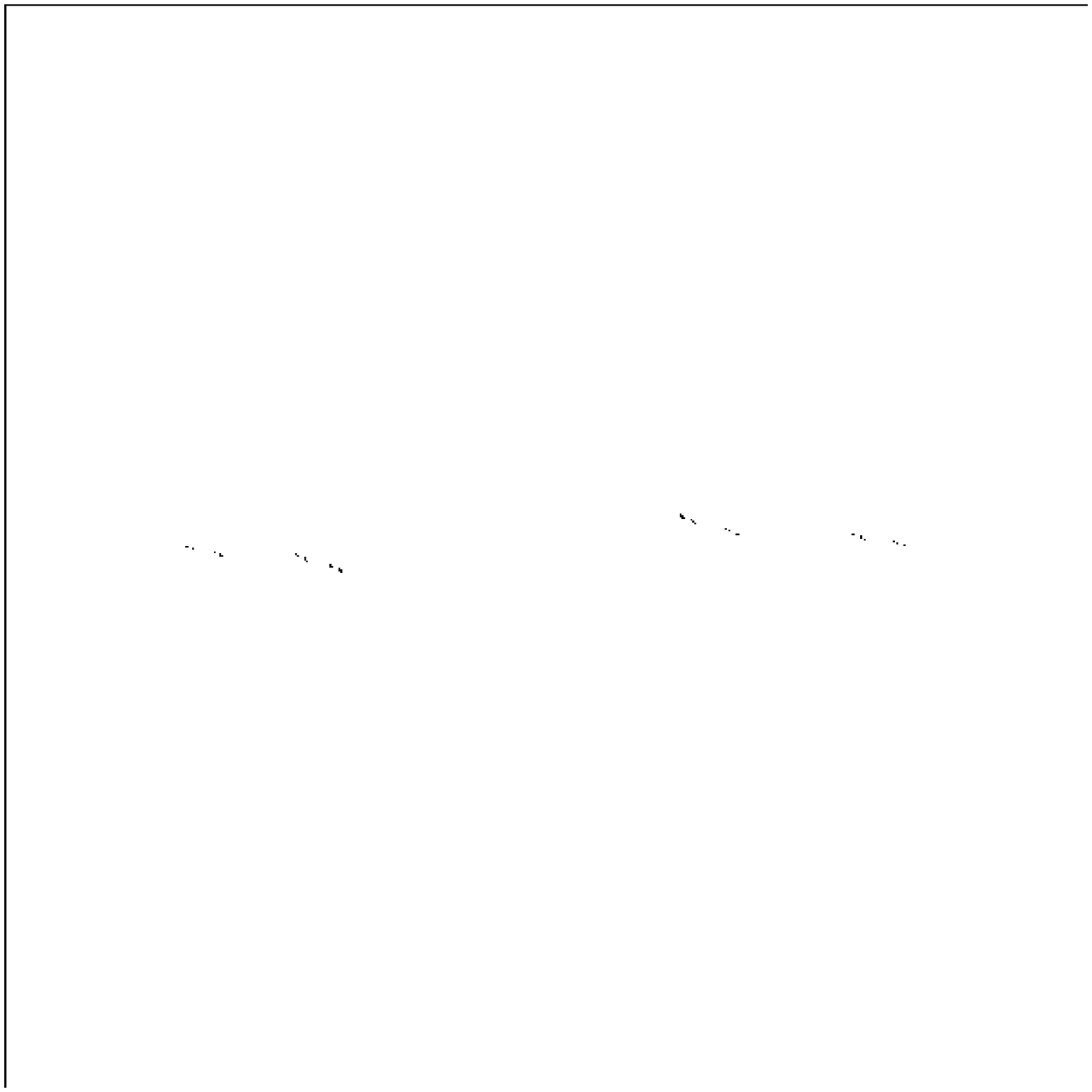}}
\put(35,70){\makebox(10,3)[c]{Fig. 3b}}
\put(10,35){\makebox(4,4)[c]{$- 4$}}
\put(67,35){\makebox(4,4)[c]{0}}
\put(35,5){\makebox(10,3)[c]{$\epsilon=-0.5$}}
\end{picture}
\begin{center}\begin{picture}(80,80)
\put(0,0){\epsfxsize=8cm\epsfbox{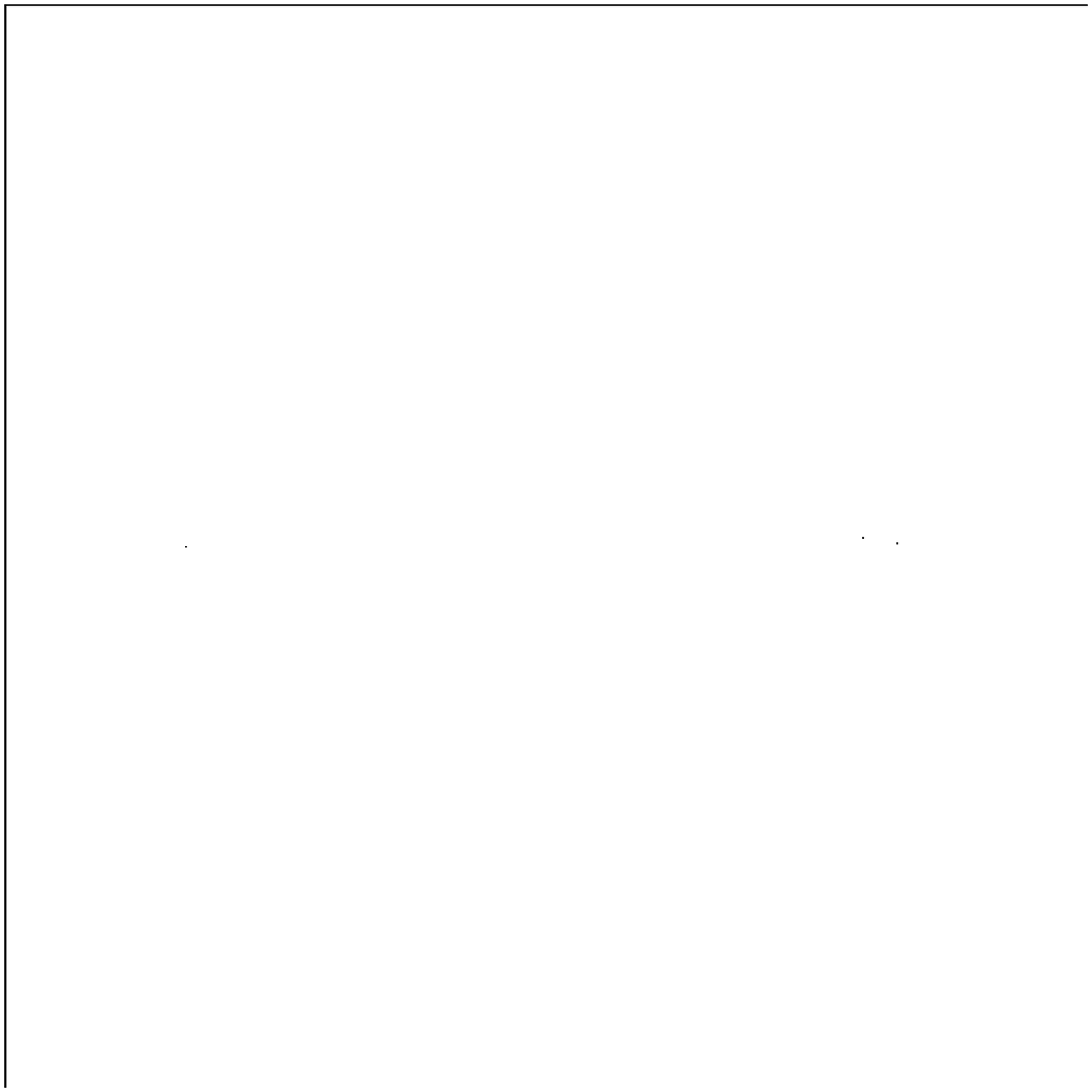}}
\put(35,70){\makebox(10,3)[c]{Fig. 3c}}
\put(10,35){\makebox(4,4)[c]{$- 4$}}
\put(67,35){\makebox(4,4)[c]{0}}
\put(35,5){\makebox(10,3)[c]{$\epsilon=0$}}
\end{picture}\end{center}
%%%%%%%%%%%%%%%%%%%%%%%%%%%%%%%%%%%%%%%%%%%%%%%%%%%%%%%%%%%%%%%%%%%%%%%%%

By studying analytical property of a piece of Toda lattice we attempted to
clarify how a nonintegrable system approaches to the integrable one. Our
argument is based on the fact that the two dimensional Toda lattice can be
disjoined into small pieces, which are integrable by themselves and are called
Toda molecules. A Toda molecule is composed from smaller pieces, which we
called Toda atoms. Hence the two dimensional Toda lattice is a crystal
consisting of Toda atoms. For such a macroscopic system being integrable every
piece must be joined very carefully not to create a Julia set.

In this onnection it will be worth while recalling that a similar property is
possessed commonly in other integrable models. In the solvable lattice models
the partition function is factorizable into a product of Boltzmann weights. The
Yang-Baxter equation is a condition imposed on the factors to be connected
properly. Another exampele is the factorizability condition imposed on the
string amplitudes which led us to the $\tau$ function of the KP hierarchy. In
any case the connection rule must be such that the symmetry characterizing the
unit blocks is preserved under the coupling.

We will be interested in studying analytically properties of the compound
system of two GLM pieces in the forth comming paper.
\ve
%%%%%%%%%%%%%%%%%%%%%%%%%%%%%%%%%%%%%%%%%

\end{document}